\documentclass[a4paper,11pt]{article}
\usepackage{pos}
\begin{document}
\title{The axion dark matter experiment MADMAX}
\author*[a]{Pascal Pralavorio}
\onbehalf{on behalf of the MADMAX Collaboration}
\affiliation[a]{Aix Marseille Univ, CNRS/IN2P3, CPPM, Marseille, France} 
\emailAdd{pralavor@cppm.in2p3.fr}
\abstract{The MAgnetized Disk and Mirror Axion eXperiment (MADMAX) is a future experiment aiming to detect dark matter axions from the galactic halo by resonant conversion to photons in a strong magnetic field. It uses a stack of dielectric disks in front of a mirror, called booster, to enhance the potential signal from axion-photon conversion over a significant mass range. Several small scale prototype systems have been developed and tested to verify the experimental principles. The current status of the experiment and its prototypes, including the ongoing research and development and remaining challenges, are presented in this contribution.}
\FullConference{ICHEP 2024\\
Prague\\
July 2024
}

\maketitle

\section{Introduction}

Axions, originally predicted to explain the absence of CP-violating effects in the strong interaction, are also prominent candidates to explain dark matter (DM) in the Universe\footnote{The term axion refers here to both QCD axions and axion-like particles not solving the strong CP problem.}. If axions explain all the DM, our galactic halo will provide axions with a density $\rho_a$ of 0.3 GeV/cm$^3$. To detect them, haloscope experiments are using an external magnetic field ($B_e$) to convert axions to photons with a coupling strength $|g_{a\gamma}|$, effectively sourcing an electric field with frequency ($\nu_a$) proportional to the axion mass ($m_a$). Since 30 years, haloscopes in the form of resonant cavities are actively searching for axions in the range 1$<\!m_a\!<$40~$\mu$eV corresponding to 0.25$<\!\nu_a\!<$10~GHz. The cavity size,  adjusted to the generated photon wavelength, decreases with the frequency giving a small conversion volume for axion with a mass above 40~$\mu$eV and thereby a reduced sensitivity. As a consequence this mass range above 40~$\mu$eV is basically uncovered even though it is very well motivated in the DM scenario in which the breaking of the Peccei-Quinn symmetry, generating the axion, happens after inflation~\cite{buschmann2022}.

\section{MADMAX Experiment}

In this context, MADMAX~\cite{MADMAX:2019pub} is proposing a detector based on the novel dielectric haloscope concept to probe this unexplored mass range. It is based on the magnetized mirror axion DM search idea~\cite{horns_2013} and uses a \textit{booster} composed of parallel dielectric disks and a mirror to enhance the axion signal. Photons, originating from axion conversion in the external magnetic field, generate coherent emissions at the dielectric disk surfaces, which can constructively interfere. A resonance enhancement is obtained when the disk spacing corresponds to about half of the photon wavelength, generating a sizeable power boost factor ($\beta^2$). This factor quantifies the signal power improvement with respect to the mirror only situation, and depends on the dielectric properties and the number of disks. This set-up allows us to decouple the conversion volume from the photon wavelength providing a possible way towards detecting axions at $m_a\!\sim\!100$~$\mu$eV~\cite{Millar:2016cjp}. Assuming a booster composed of 80 dielectric disks, of $A$=1~m$^2$ surface, housed in a 4~K cryostat inserted in a 10~T dipole magnet, a QCD axion with a coupling strength $|g_{a\gamma}|\simeq 2 \cdot 10^{-14} \ \textrm{GeV}^{-1}$ seems to be within reach for detection with a signal-to-noise ratio (SNR) of 5 after few days ($\Delta t$):
\begin{equation}
|g_{a\gamma}| \simeq 2 \cdot 10^{-14} \ \textrm{GeV}^{-1} 
\left(\frac{m_a}{100 \ \mu\textrm{eV}}\right)^{5/4} 
\left(\frac{10 \ \textrm{T}}{B_e}\right) 
\left(\frac{5\cdot 10^4}{\beta^2} \cdot 
\frac{1 \ \textrm{m}^2}{A} \cdot
\frac{\textrm{SNR}}{5} \cdot
\frac{T_{sys}}{4 \ \textrm{K}}\right)^{1/2}  
\left(\frac{\textrm{1.8 d}}{\Delta t}\right)^{1/4} 
\end{equation}
Axion mass can be scanned by moving discs with piezoelectric motors, needed to control the disk position at O($\mu$m) level. All these constraints guide the final design of the detector, planned to be located at DESY in the H1 iron yoke. The main challenges involve building a O(10)~T dipole magnet, developing a receiver chain in the O(10)~GHz frequency regime and operating the booster with O(1)~m diameter disks at 4~K under a 10~T magnetic field. 

Since 2020 an intense prototyping phase has started to validate the concept and gradually build the final booster. The idea is to operate small-scale prototype boosters composed of 20 or 30~cm diameter sapphire disks to validate the dielectric haloscope concept and to perform first dark matter searches in an uncharted phase space as detailed in the next section.

\section{First Dark matter searches with a dielectric haloscope}
\begin{figure}[t]
\includegraphics[width=0.95\linewidth]{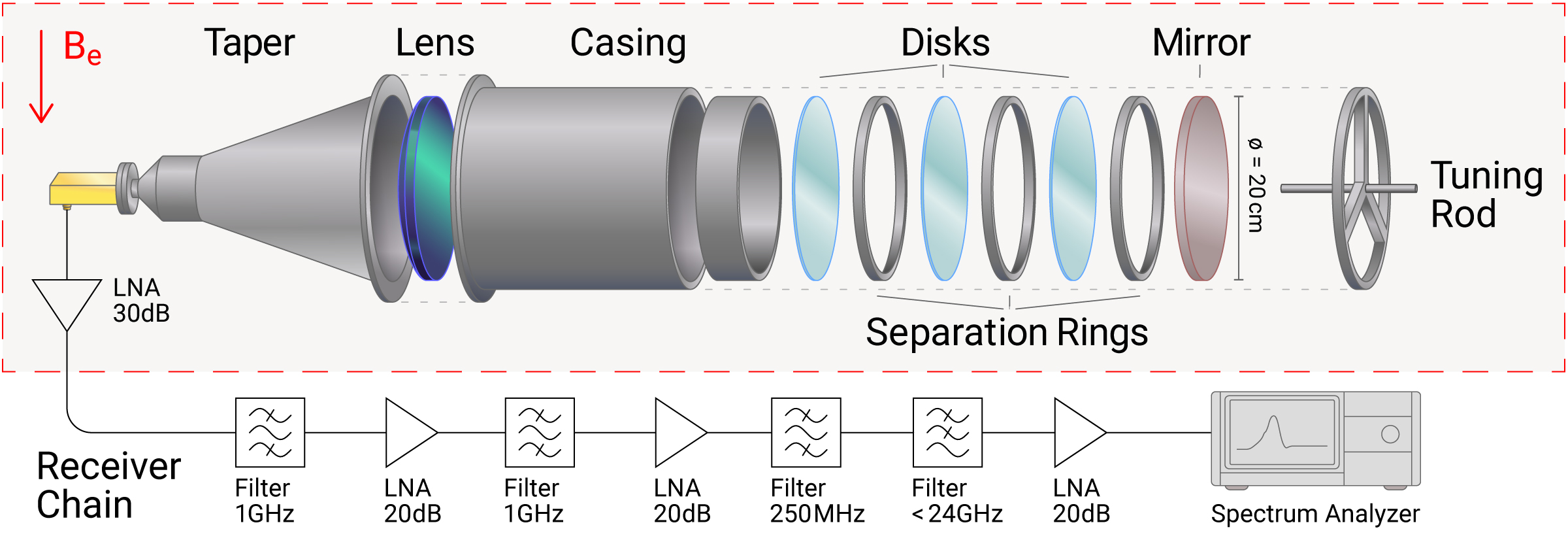}
\caption{Sketch of the closed MADMAX booster prototype with 3 disks of 20~cm diameter and a mirror inserted in a casing~\cite{Garcia:2024xzc}. The heterodyne receiver system, located outside the magnetic field $B_e$ is composed of low noise amplifiers (LNAs) and filters coupled to a commercial real-time spectrum analyser.}
\label{fig:Prototype}
\end{figure}

Following the acceptation by the CERN Research Board in 2020, the MADMAX prototypes are tested during the beam shutdown period in a dipole magnet of 1.6~T located in the North Area. After the magnet area refurbishment in 2021, the tests performed in 2022 and 2023 demonstrated that the radio frequency environment is suitable for MADMAX and that the livetime is inversely proportional to square of the data taking noise according to the Dicke's radiometer equation. During these tests, a Y-factor method was also developed to calibrate the receiver chain power and extract the system temperature ($T_{sys}$). During winter 2024, 14.5 days of data were taken with the closed booster prototype shown in Figure~\ref{fig:Prototype} -- closed since disks are located in a casing. In this set-up it is possible to adjust the inter-disk spacing by changing the separation rings and obtain 2 configurations centered on 18.55 and 19.21~GHz, corresponding to 76.72 and 79.45~$\mu$eV axion masses. A micrometer change in separation between the mirror and the closest disk using the tuning rod (Figure~\ref{fig:Prototype}) allows us to further change the booster peak by O(10)~MHz, resulting in five different configurations. 

\begin{figure}[htbp]
\includegraphics[width=0.575\linewidth]{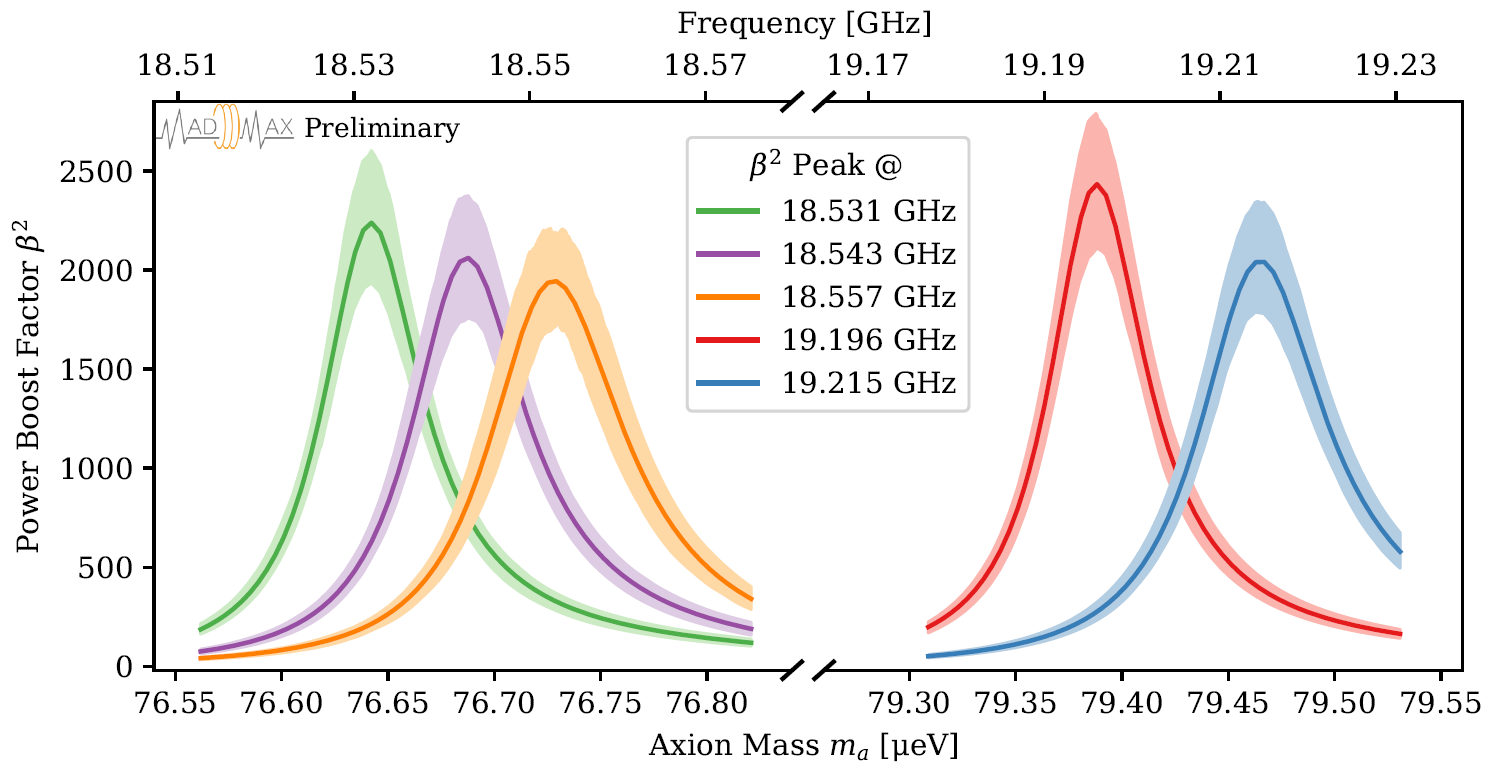}
\includegraphics[width=0.415\linewidth]{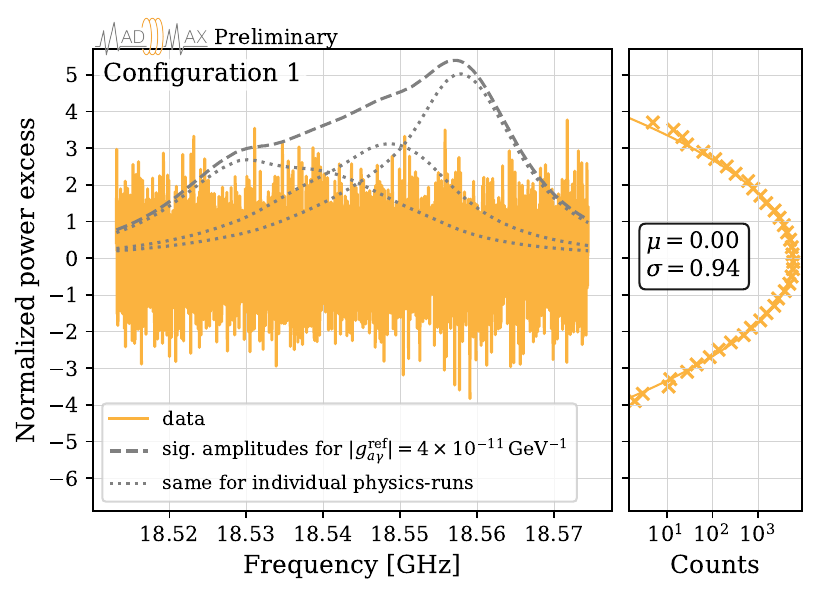}
\centering
\includegraphics[width=0.65\linewidth]{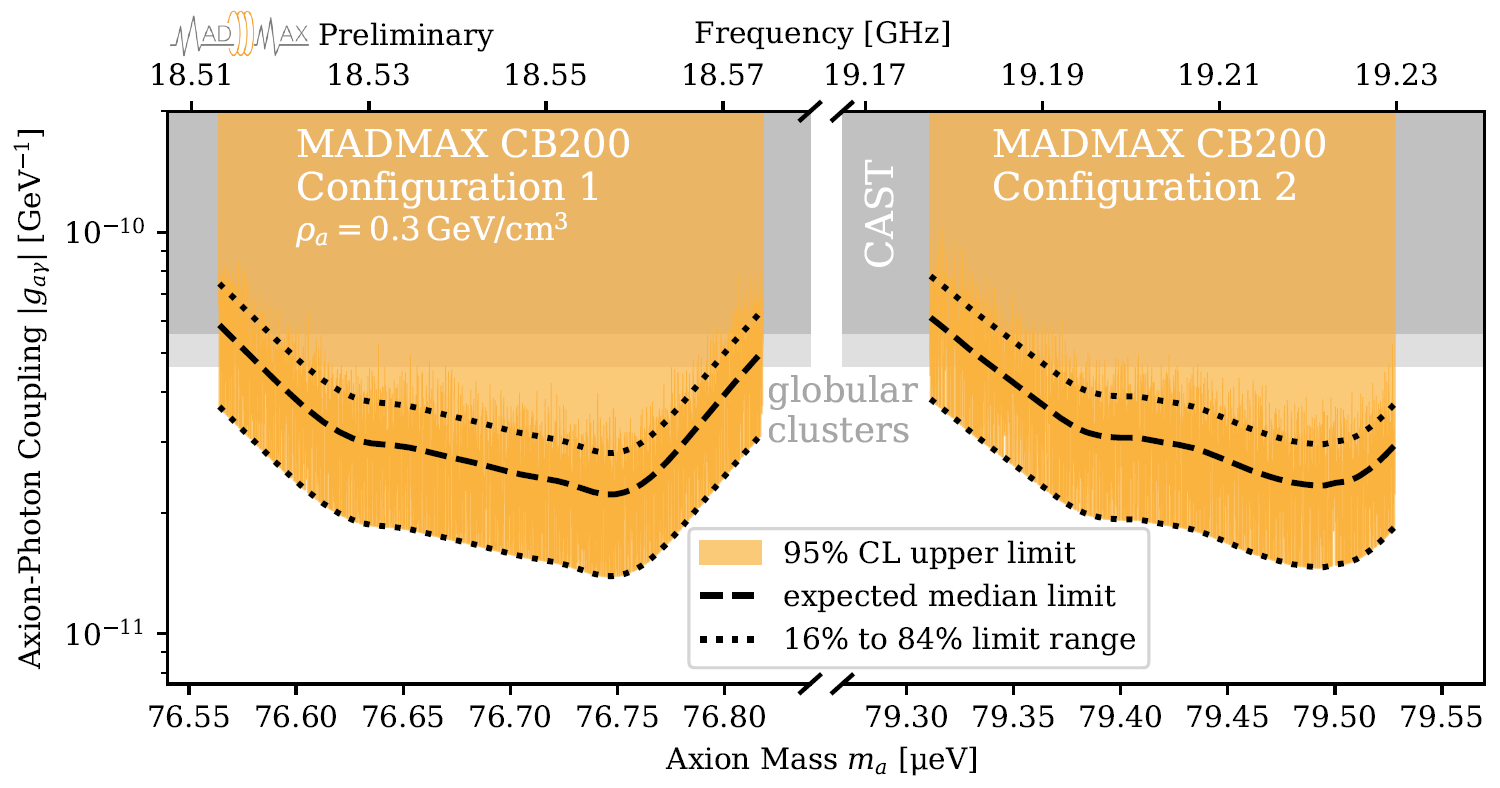}
\caption{Results of the closed booster axion DM search~\cite{Garcia:2024xzc}. Top left: Boost factor for the 5 configurations. Top right: Grand spectrum around 18.55~GHz. Dashed grey lines indicate the envelope of expected amplitude in each frequency bin of a potential signal. Bottom: 95\% CL exclusion limit in the $|g_{a\gamma}|-m_a$ plane.}
\label{fig:Axion}
\end{figure}

The boost factor, defined for an axion induced signal, is determined using the strong correlation between reflectivity measurement of the system and a 1D-model implemented in ADS software~\cite{Garcia:2024xzc}. The peak values of $\beta^2$ are around 2000 with uncertainties around 15\% (Figure~\ref{fig:Axion} top left). The data analysis procedure followed that of HAYSTACK~\cite{Brubaker:2017rna}. It combines a set of individual power spectra to a grand spectrum, filtered using a Savistky-Golay function. In absence of an axion signal, the grand spectrum is expected to be a Gaussian centered on 0 with a width of 0.94~\cite{Brubaker:2017rna}. As shown in Figure~\ref{fig:Axion} top right, no signal is visible. An upper limit is extracted in the $|g_{a\gamma}|$-$m_a$ plane (Figure~\ref{fig:Axion} bottom), exceeding previous world-best limits from the CAST helioscope and astrophysical indirect constraints in a mass range of 0.40~$\mu$eV around 78~$\mu$eV~\cite{Garcia:2024xzc}. This result validates for the first time the dielectric haloscope concept in a magnetic field. 

\begin{figure}[htbp]
\includegraphics[width=0.485\linewidth]{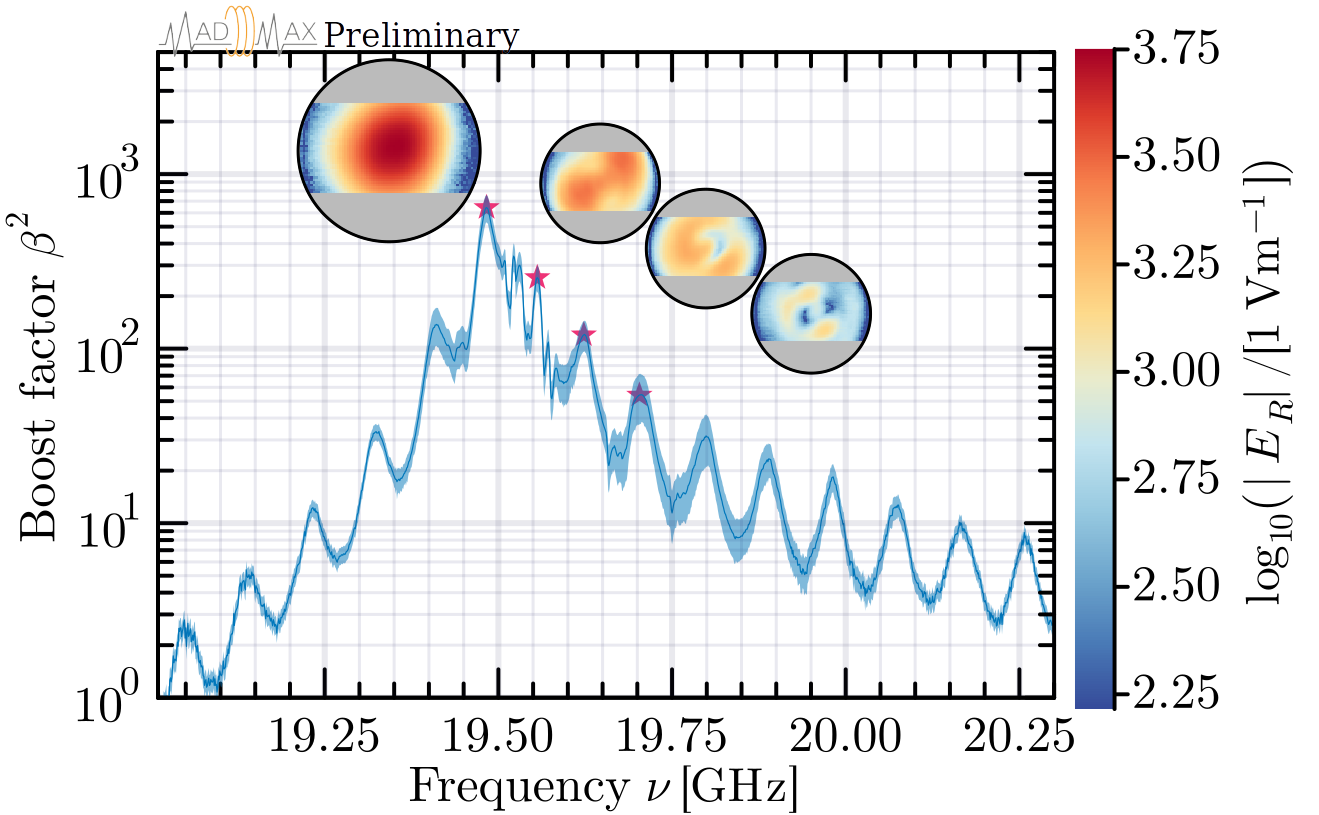}
\includegraphics[width=0.490\linewidth]{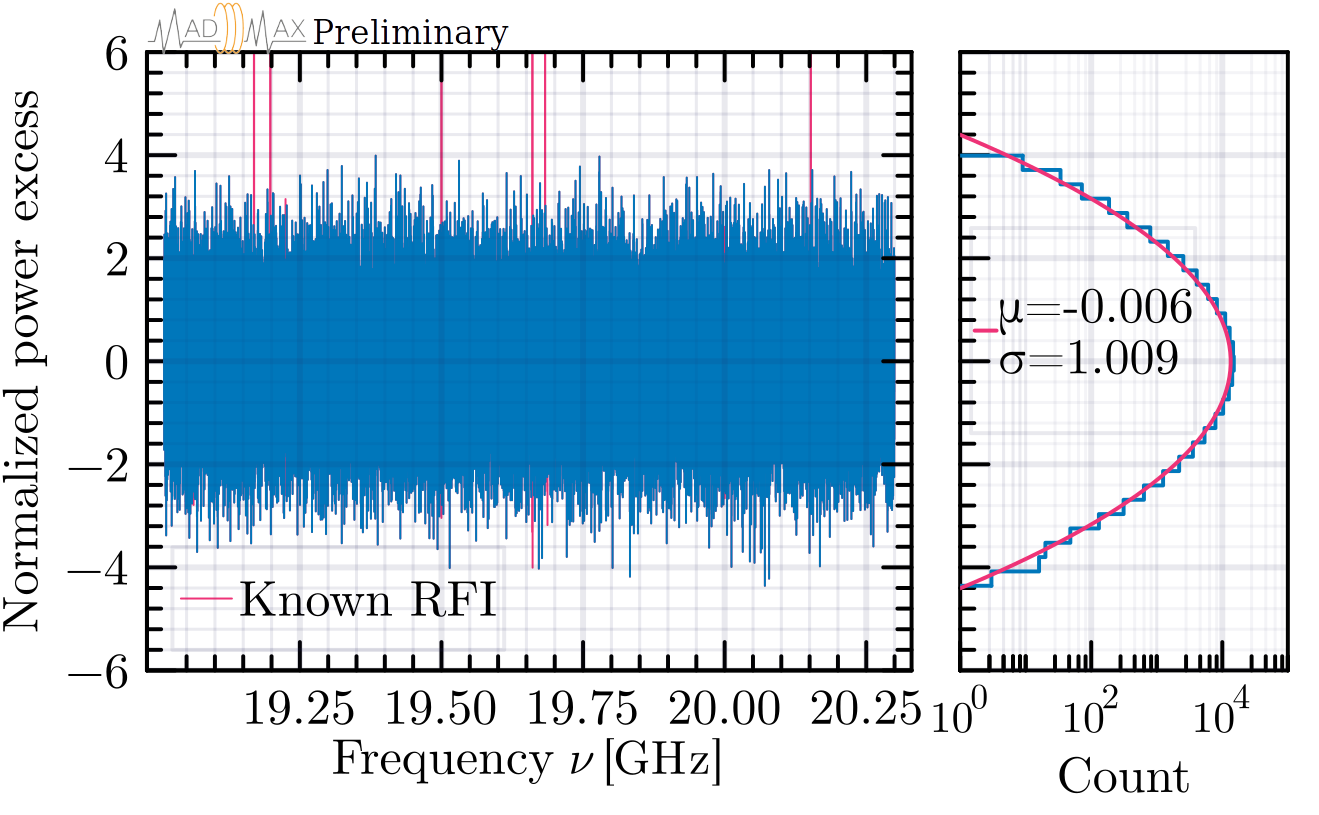}
\centering
\includegraphics[width=0.65\linewidth]{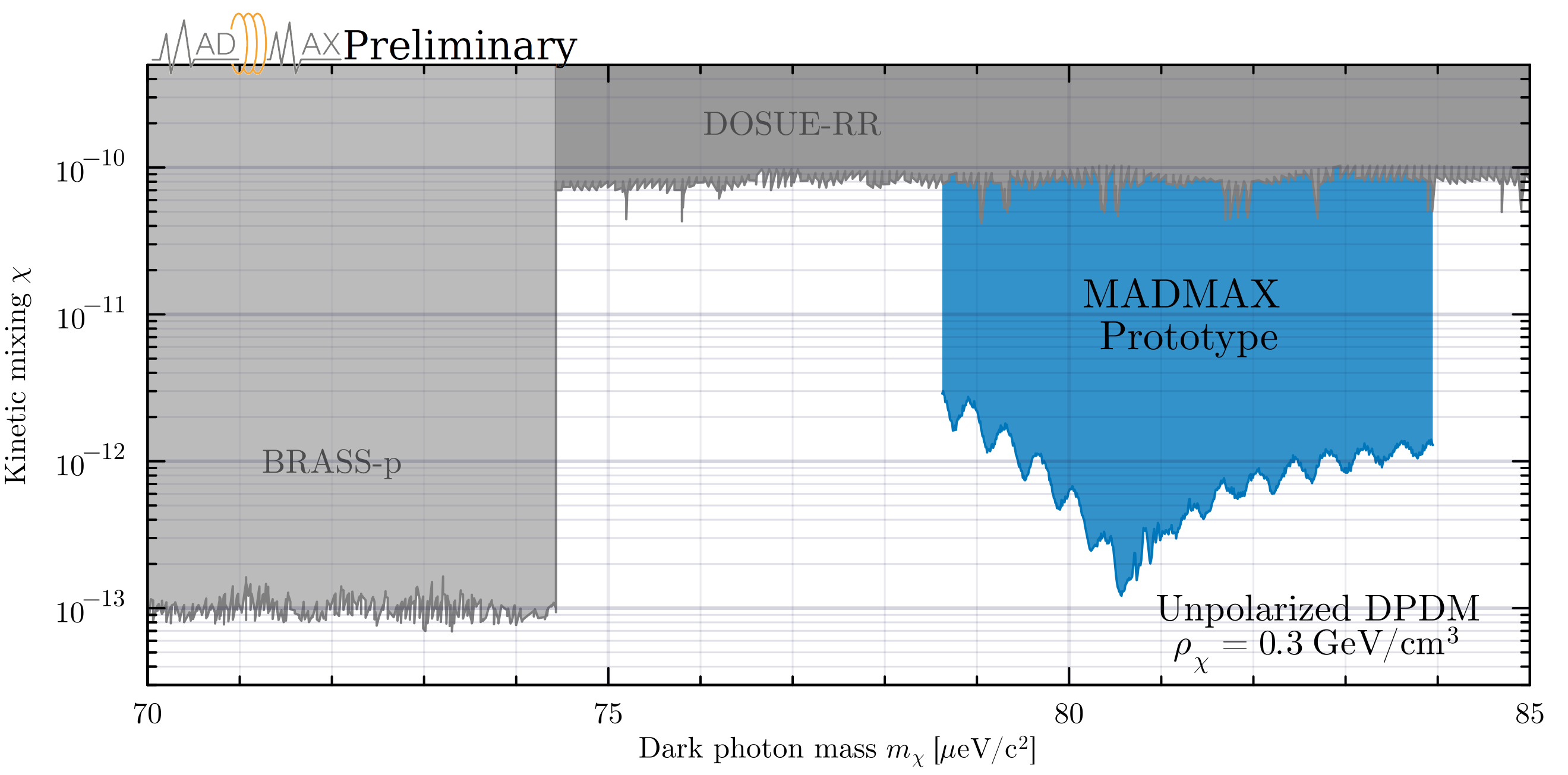}
\caption{Results of the open booster dark photon DM search~\cite{MADMAX:2024jnp}. Top left: Boost factor measurement. Top right: Grand spectrum around 19.5~GHz. Bottom: 95\% CL exclusion limit in the $\chi-m_{\chi}$ where $\chi$ is the mixing parameter between photon and dark photon.}
\label{fig:DarkPhoton}
\end{figure}

The next step is to operate an open booster, i.e. a booster without the casing shown in Figure~\ref{fig:Prototype}. This prototype is equipped with 3 fixed disks and a mirror of 30~cm diameter and operate at room temperature without magnetic field in a shielded laboratory. A dedicated bead-pull method~\cite{Egge:2022gfp,Egge:2023cos} enables the in-situ measurement of the boost factor from the electromagnetic response of the booster: a small bead is introduced in the booster and a reflection-induced electric field excited by a Vector Network Analyzer is inferred from small changes in the booster reflection coefficient at the bead’s position. The boost factor distribution is shown in Figure~\ref{fig:DarkPhoton} top left, where the insets show the transverse electric field between the mirror and the first disk at the indicated frequencies (stars). A mix of higher-order transverse modes can also resonate causing the additional smaller peaks in the $\beta^2$ distribution. A $\beta^2$ peak of around 600 is obtained with an uncertainty around 15\%. Using this open booster set-up, a dark photon DM ($\chi$) search was performed during 12 days. Figure~\ref{fig:DarkPhoton} top right shows the grand spectrum where no signal of unknown origin is observed. Figure~\ref{fig:DarkPhoton} bottom shows the corresponding 95\% CL exclusion limits in the $\chi-m_{\chi}$ plane. MADMAX set the world best limits in a large 78.6-83.9 $\mu$eV mass range, 1 to 3 orders of magnitude below the previous limits~\cite{MADMAX:2024jnp}, assuming a local dark photon DM density of $\rho_\chi = 0.3$ GeV/cm$^3$. This result demonstrates the broadband capacity of the MADMAX booster.

\begin{figure}[htbp]
\includegraphics[width=0.485\linewidth]{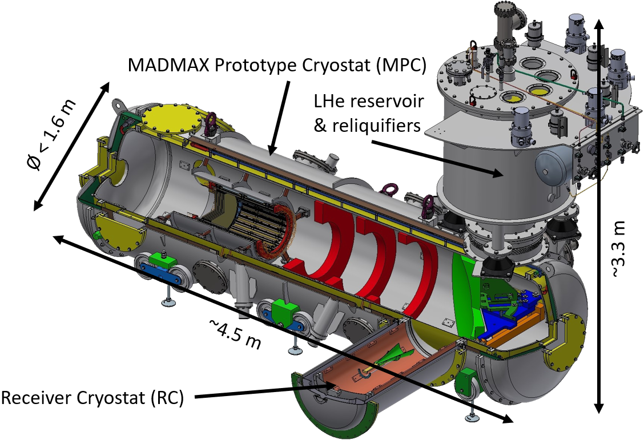}
\includegraphics[width=0.485\linewidth]{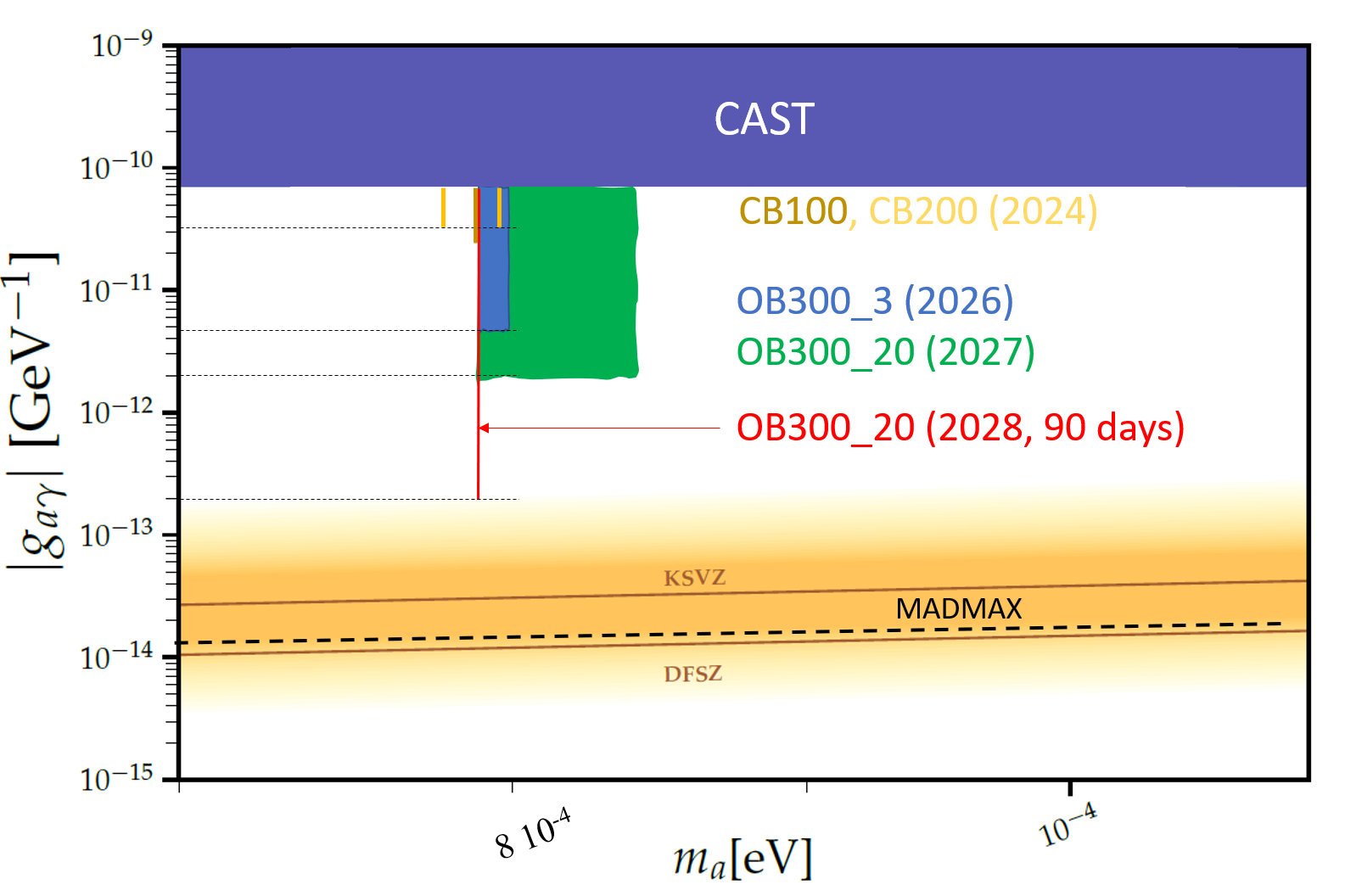}
\caption{Left: Sketch of the MADMAX prototype cryostat. Right: Physics reach of the MADMAX closed booster prototypes (CB100 and CB200) in 2024 as well as the final open booster (OB300) in 2026-28.}
\label{fig:Future}
\end{figure} 

\section{Towards the final MADMAX experiment}

To demonstrate the full scanning capacity of MADMAX an open booster with moveable disks must be considered. In this booster, the disks are moved by piezoelectric motors~\cite{Garutti:2023stk} which slide on three parallel rails hold in a backbone structure. Such a set-up was successfully tested both inside a magnetic field of 1.6 T and at cryogenic temperatures down to 35K. The measurements of the velocity and positioning accuracy of the disk are found to match the MADMAX requirements~\cite{MADMAX:2024pil}. The final step of the prototype program is therefore to perform long runs with an open booster composed of 3 to 20 movable disks of 30~cm diameter in the CERN dipole magnet during the LHC shutdown in 2026-29. The booster will be placed in a stainless steel cryostat (Figure~\ref{fig:Future} left) presently in construction and to be delivered in 2025 on the DESY campus. Such an apparatus, close to the final design, could be operated 3 months per year enabling to increase significantly the mass coverage and the gain in sensitivity by up to 2 orders of magnitude (Figure~\ref{fig:Future} right). 

In parallel, the work on the final magnet is also on-going: the design of the 18 skateboard coils with the novel copper CICC conductor has been completed in 2021 and a dedicated set-up has demonstrated that these coils are safe in terms of quench protection~\cite{macqu}. A demonstrator coil is presently in construction. Similarly, the receiver chain, which currently use High Electron Mobile Transistor for the LNA, plan to upgrade to Josephson junction (TWPA) to further reduce the noise.

\section{Conclusion}

The MADMAX collaboration is proposing to use for the first time a dielectric haloscope to scan QCD axion mass around 100~$\mu$eV. It is presently in a vivid prototyping phase to demonstrate the capacity of this new concept. Recent achievements with booster prototypes enabled to validate the mechanics of the booster at cryogenic temperature and under magnetic field, to establish a method to measure the boost factor in-situ while performing DM searches in the 18-20~GHz frequency regime, demonstrating the power and versatility of the dielectric haloscope concept. 

\bibliographystyle{JHEP}
\bibliography{main}
\end{document}